\documentclass[sigconf]{acmart}

\AtBeginDocument{%
  }

\usepackage{enumitem}   
\usepackage{wasysym}    
\newcommand{\fc}{\CIRCLE}      
\newcommand{\hc}{\LEFTcircle}  
\newcommand{\ec}{\Circle}      

\usepackage{tikz}
\usetikzlibrary{positioning, arrows.meta, fit, backgrounds}
\usepackage{forest}     
\usepackage{fontawesome5} 

\definecolor{cAgent}{HTML}{3B4CC0}
\definecolor{cDB}{HTML}{2A7FB8}
\definecolor{cFile}{HTML}{138D75}
\definecolor{cRAG}{HTML}{8E44AD}
\definecolor{cTool}{HTML}{B5650C}
\definecolor{cPeer}{HTML}{27744E}
\definecolor{cMem}{HTML}{B03A2E}
\definecolor{cLeak}{HTML}{E74C3C}
\definecolor{cUser}{HTML}{6B7280}
\newcommand{\leakicon}{\textcolor{cLeak}{\small\faExclamationTriangle}}

\setcopyright{none}            
\copyrightyear{2026}
\acmYear{2026}
\acmConference[Preprint]{Working draft}{2026}{}
\renewcommand\footnotetextcopyrightpermission[1]{} 
\begin{document}

\title{Agents That Know Too Much: A Data-Centric Survey of Privacy in
  LLM Agents}

\author{Nada Lahjouji}
\email{nlahjouj@uci.edu}
\affiliation{%
  \institution{University of California, Irvine}
  \city{Irvine}
  \state{California}
  \country{USA}
}

\author{Ashwin Gerard Colaco}
\email{acolaco@uci.edu}
\affiliation{%
  \institution{University of California, Irvine}
  \city{Irvine}
  \state{California}
  \country{USA}
}
\renewcommand{\shortauthors}{Lahjouji and Colaco}

\begin{abstract}
  Large language model agents increasingly query databases, search
  document collections, call external APIs, remember past interactions,
  and act on a user's behalf. As they move from answering questions to
  operating over sensitive data, privacy becomes harder to enforce. An
  agent touches many data sources, runs multi-step workflows, keeps
  state across sessions, and acts with delegated permissions. Sensitive
  information can therefore leak not only through its final answer but
  through the queries it issues, the intermediate results it handles,
  the memory it writes, and the messages it exchanges with other agents.
  We survey the privacy of LLM agents from a \emph{data-centric} view,
  organizing the field around the data an agent touches rather than by
  attack type, and we use \emph{data agent} as shorthand for an LLM
  agent that works with data. Research on these risks is active but
  scattered across retrieval-augmented generation, text-to-SQL
  interfaces, agent memory, prompt injection, access control, and
  contextual privacy. This survey brings that work together: we
  taxonomize the data sources an agent touches, the privacy risks each
  source creates, and the governance mechanisms that address them; we
  map the benchmarks used to measure these risks and identify what is
  missing; and we set out the open problems. Two findings recur: among
  governance mechanisms only information-flow control covers both
  compositional and cross-session inference leakage, the two
  least-protected risks; and no benchmark drives an agent across its
  data surfaces under one privacy policy, the instrument the field most
  lacks. Our goal is a reference that situates the scattered literature
  and gives future work a common framing.
\end{abstract}



\maketitle
\pagestyle{plain}   

\section{Introduction}
\label{sec:intro}

For most of their short history, large language models were used as
chatbots: a user typed a question, the model produced an answer, and the
interaction ended. Privacy questions in that setting concerned what the
model had memorized from its training corpus and whether it would repeat
it~\cite{carlini2021extracting, huang2022pii, lukas2023pii, nasr2023scalable}. The interaction
left no state, touched no external data, and took no action in the world.

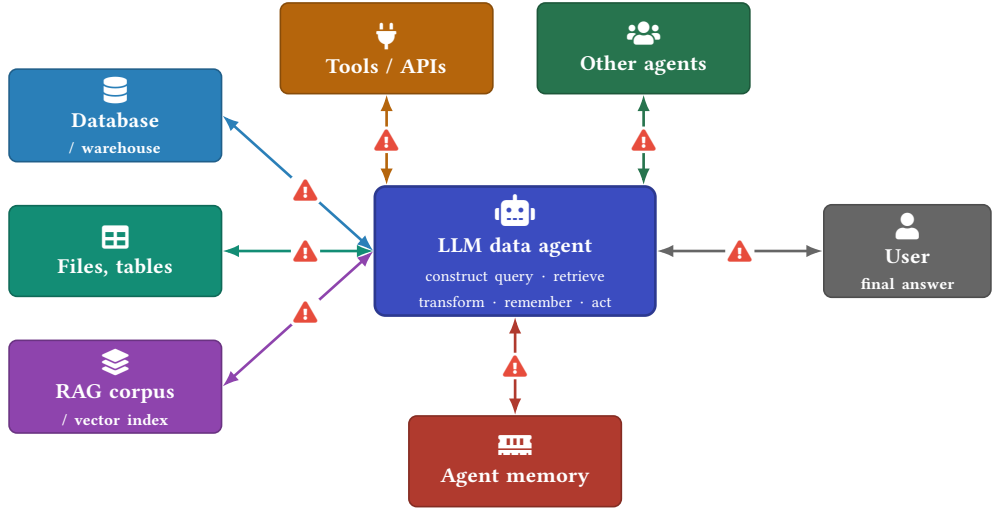
\begin{figure*}[t]
\centering
\begin{tikzpicture}[
  font=\small,
  surf/.style={draw=#1!75!black, line width=0.6pt, rounded corners=3pt, fill=#1,
    text=white, align=center, minimum height=12mm, text width=26mm, inner sep=3pt},
  agent/.style={draw=cAgent!75!black, line width=1pt, rounded corners=4pt, fill=cAgent,
    text=white, align=center, minimum height=17mm, text width=35mm, inner sep=3pt},
  io/.style={draw=black!70, line width=0.6pt, rounded corners=3pt, fill=black!60,
    text=white, align=center, minimum height=12mm, text width=20mm, inner sep=3pt},
  ch/.style={{Latex[length=2.2mm]}-{Latex[length=2.2mm]}, line width=1pt, draw=#1},
  chout/.style={-{Latex[length=2.4mm]}, line width=1pt, draw=#1},
  leak/.style={fill=white, inner sep=1pt, rounded corners=1pt},
]
\node[agent] (ag) {{\LARGE\faRobot}\\[1pt]\textbf{LLM data agent}\\[1pt]
  {\scriptsize construct query $\cdot$ retrieve}\\{\scriptsize transform $\cdot$ remember $\cdot$ act}};
\node[surf=cDB, above left=3mm and 20mm of ag] (db) {{\large\faDatabase}\\\textbf{\small Database}\\\textbf{\scriptsize / warehouse}};
\node[surf=cFile, left=20mm of ag] (fl) {{\large\faTable}\\\textbf{\small Files, tables}};
\node[surf=cRAG, below left=3mm and 20mm of ag] (rag) {{\large\faLayerGroup}\\\textbf{\small RAG corpus}\\\textbf{\scriptsize / vector index}};
\node[surf=cTool, above=12mm of ag, xshift=-17mm] (tools) {{\large\faPlug}\\\textbf{\small Tools / APIs}};
\node[surf=cPeer, above=12mm of ag, xshift=17mm] (peers) {{\large\faUsers}\\\textbf{\small Other agents}};
\node[surf=cMem, below=13mm of ag] (mem) {{\large\faMemory}\\\textbf{\small Agent memory}};
\node[io, right=22mm of ag] (user) {{\large\faUser}\\\textbf{\small User}\\\textbf{\scriptsize final answer}};
\draw[ch=cDB] (db.east) to node[leak, pos=0.55]{\leakicon} (ag.west);
\draw[ch=cFile] (fl.east) to node[leak, pos=0.55]{\leakicon} (ag.west);
\draw[ch=cRAG] (rag.east) to node[leak, pos=0.55]{\leakicon} (ag.west);
\draw[ch=cTool] ([xshift=-17mm]ag.north) to node[leak, pos=0.5]{\leakicon} (tools.south);
\draw[ch=cPeer] ([xshift=17mm]ag.north) to node[leak, pos=0.5]{\leakicon} (peers.south);
\draw[ch=cMem] (ag.south) to node[leak, pos=0.5]{\leakicon} (mem.north);
\draw[ch=black!60] (ag.east) to node[leak, pos=0.5]{\leakicon} (user.west);
\end{tikzpicture}
\caption{A data agent and the surfaces it operates over. The agent constructs
queries to and receives results from heterogeneous data surfaces (databases and
warehouses, files, and RAG corpora), calls tools and other agents, and writes to
persistent memory. Each \leakicon\ marks a channel on which sensitive information
can leak: the queries the agent constructs, the results it retrieves, the
arguments it passes to tools, the messages it exchanges with other agents, the
state it writes to memory, and the final answer. Privacy must therefore be
governed across the whole execution, not only at the output.}
\label{fig:overview}
\end{figure*}

This setting no longer describes how language models are deployed.
Equipped with the ability to reason about a goal, call tools, and observe
the results~\cite{yao2023react, schick2023toolformer}, language models now
operate as \emph{agents} that pursue multi-step tasks with little
supervision, carry memory across turns and sessions~\cite{park2023generative},
and coordinate with other agents~\cite{wu2023autogen, han2024llmmas}. A large
share of these agents exist to work with data. They translate a request into
a database query, retrieve passages from a document collection, join results
across sources, summarize what they find, write intermediate findings to
memory, and act on the result. We refer to such systems as \emph{data
agents}: LLM-driven agents whose purpose is to query, retrieve, transform,
remember, and act on data on a user's behalf~\cite{luo2026dataagents,
tang2025dataanalyst, fu2025autonomousdataagents}. Data agents already drive
enterprise analytics over warehouses~\cite{sun2025agenticdata, lei2024spider2},
scientific and data-science question answering~\cite{zhang2025deepanalyze}, and
personal assistants that read a user's files, mail, and calendar.

The data these agents touch is sensitive. A clinical agent reads patient
records, an enterprise analytics agent reads payroll and customer tables, and
a personal assistant reads private correspondence. When a model only answered
questions, privacy could be enforced at one place, the final answer, by
filtering what the model was about to say. A data agent eliminates this single enforcement point for five reasons. First, it touches \emph{many heterogeneous data
sources} at once: relational databases, vector indexes built over private
documents, files, external APIs, and its own memory, each exposing data in different ways (Figure~\ref{fig:overview}). Second, it runs \emph{multi-step workflows} in which sensitive
information flows through intermediate artifacts long before any answer is
produced: the generated query, the retrieved rows, the arguments passed to a
tool, and the text written to a scratchpad, each of which can expose sensitive information~\cite{elyagoubi2026agentleak, alizadeh2025simplepi}. Third, it keeps
\emph{persistent state}: data written to memory in one session can surface in
another, to a different user~\cite{wang2025mextra, chen2024agentpoison}.
Fourth, it acts with \emph{delegated authority}, often holding broader
permissions than any single task requires, so a compromised or misdirected
agent can reach far more data than its current request~\cite{ji2026privilege,
li2025accesscontrol}. Fifth, every protective transformation trades against
\emph{utility}: the more an agent redacts, abstracts, or withholds, the less
useful its answers become, and that tension must be managed rather than
resolved once~\cite{abedini2025masksql, koga2024dprag}. Privacy for data
agents is therefore not a property of a single output. It is a property of an
execution that spans data sources, intermediate channels, memory, and actions.

Each of these risks has an established literature, but these literatures have
developed largely in isolation, in research communities with little overlap.
Work on natural language interfaces to databases measures and
defends against sensitive data leaking through generated
SQL~\cite{song2024securesql, abedini2025masksql, liu2025safenlidb}. Work on
retrieval-augmented generation shows that private corpora can be extracted or
have their membership inferred through an agent's
queries~\cite{zeng2024goodbad, anderson2025ragmia, qi2025spillbeans}. Work on
agent memory demonstrates extraction and poisoning of stored
state~\cite{wang2025mextra, chen2024agentpoison, dong2025minja}. Work on
prompt injection shows that untrusted data and tools can hijack an agent into
exfiltrating what it sees~\cite{greshake2023indirect, zhan2024injecagent,
debenedetti2024agentdojo}. A separate line builds access control and
information-flow control for agents~\cite{costa2025fides, debenedetti2025camel,
shi2025progent}, and another studies whether agents respect contextual norms
of appropriate information flow~\cite{mireshghallah2024confaide,
bagdasarian2024airgap, shao2024privacylens}. None of them organizes its findings around the data the agent handles,
and so their combined account of an end-to-end data agent remains
fragmented.

We unify these lines under a single, data-centric view of LLM agents. Rather than
organize by attack type, as broad agent-security surveys
do~\cite{he2024emerged, gan2024navigating, qi2026trustworthy}, we organize
around the data itself: where it resides, the privacy risks it incurs, and how to
govern it. Concretely, we make the following contributions.

\begin{itemize}[leftmargin=*]
\item We motivate a data-centric view of agent privacy and explain why an LLM
agent that works with data, which we call a data agent, has privacy properties
different from those of stateless chatbots and generic tool-using agents
(Sections~\ref{sec:intro} and~\ref{sec:background}).

\item We give three taxonomies and the tables that connect them: the
\emph{data surfaces} a data agent touches (Section~\ref{sec:surfaces}), the
\emph{privacy risks} each surface creates (Section~\ref{sec:risks}), and the
\emph{governance mechanisms} that address them (Section~\ref{sec:governance}).
Three tables connect the taxonomies: surfaces to the disclosure outcomes they
create (Table~\ref{tab:surfacerisk}), outcomes to the vectors that trigger them
(Table~\ref{tab:outcomevector}), and outcomes to the controls that address them
(Table~\ref{tab:matrix}). They are intended as a reusable reference for both
system builders and reviewers.

\item We map the benchmarks that measure these risks, compare what each one
covers, and identify the gap that no current benchmark fills: an end-to-end
data-agent workflow evaluated under an explicit privacy policy
(Section~\ref{sec:evaluation}).

\item We survey where data agents are deployed (Section~\ref{sec:applications})
and set out the open problems for the field (Section~\ref{sec:open}).
\end{itemize}

Two practical conclusions recur. A builder who can adopt only one advanced
mechanism should adopt information-flow control, the single mechanism in our map
that covers both compositional leakage and inference across a session. And the
instrument the field most lacks is an end-to-end benchmark that drives an agent
across its data surfaces under one explicit privacy policy; we argue for it in
Section~\ref{sec:evaluation}.

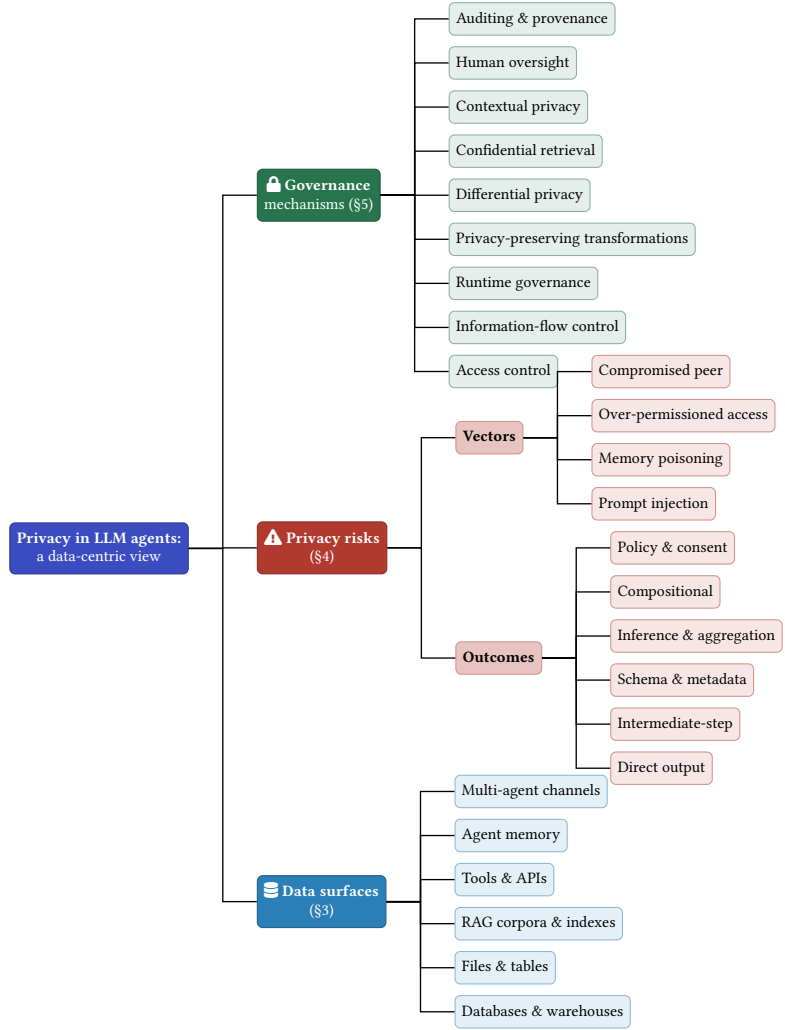
\begin{figure*}[t]
\centering
\begin{forest}
  for tree={
    grow=east,
    parent anchor=east,
    child anchor=west,
    draw, rounded corners=2pt,
    align=center,
    anchor=west,
    font=\scriptsize,
    l sep=9mm,
    s sep=1.4mm,
    inner sep=2.6pt,
    fill=gray!8,
    edge={line width=0.5pt},
    edge path={\noexpand\path[\forestoption{edge}] (!u.parent anchor) -- ++(4.5mm,0) |- (.child anchor)\forestoption{edge label};},
  }
  [{\textbf{Privacy in LLM agents:}\\a data-centric view}, fill=cAgent, text=white, draw=cAgent!75!black
    [{\faDatabase~\textbf{Data surfaces}\\(\S\ref{sec:surfaces})}, fill=cDB, text=white, draw=cDB!75!black,
       for children={fill=cDB!12, draw=cDB!55, text=black}
      [Databases \& warehouses]
      [Files \& tables]
      [RAG corpora \& indexes]
      [Tools \& APIs]
      [Agent memory]
      [Multi-agent channels]
    ]
    [{\faExclamationTriangle~\textbf{Privacy risks}\\(\S\ref{sec:risks})}, fill=cMem, text=white, draw=cMem!75!black
      [{\textbf{Outcomes}}, fill=cMem!30, draw=cMem!55, text=black,
         for children={fill=cMem!12, draw=cMem!55, text=black}
        [Direct output]
        [Intermediate-step]
        [Schema \& metadata]
        [Inference \& aggregation]
        [Compositional]
        [Policy \& consent]
      ]
      [{\textbf{Vectors}}, fill=cMem!30, draw=cMem!55, text=black,
         for children={fill=cMem!12, draw=cMem!55, text=black}
        [Prompt injection]
        [Memory poisoning]
        [Over-permissioned access]
        [Compromised peer]
      ]
    ]
    [{\faLock~\textbf{Governance}\\mechanisms (\S\ref{sec:governance})}, fill=cPeer, text=white, draw=cPeer!75!black,
       for children={fill=cPeer!12, draw=cPeer!55, text=black}
      [Access control]
      [Information-flow control]
      [Runtime governance]
      [Privacy-preserving transformations]
      [Differential privacy]
      [Confidential retrieval]
      [Contextual privacy]
      [Human oversight]
      [Auditing \& provenance]
    ]
  ]
\end{forest}
\caption{Organization of the survey. We take a data-centric view of privacy in
LLM agents, structured around the data an agent touches: the data surfaces it
operates over (Section~\ref{sec:surfaces}), the privacy risks each surface
creates (Section~\ref{sec:risks}), and the governance mechanisms that address
them (Section~\ref{sec:governance}). Cross-tables connect them, mapping surfaces
to risks (Table~\ref{tab:surfacerisk}), outcomes to the vectors that trigger them
(Table~\ref{tab:outcomevector}), and risks to controls (Table~\ref{tab:matrix}).}
\label{fig:taxonomy}
\end{figure*}

Several recent surveys treat the security and privacy of LLM agents broadly and
organize by attack type~\cite{he2024emerged, gan2024navigating,
qi2026trustworthy, ling2026towardsecure}. Two works share our premise that agent
privacy extends across the whole execution rather than the final output: a
position paper that reframes privacy beyond training-data
memorization~\cite{mireshghallah2025notmemorization}, and a survey that
decomposes agentic data leakage by architectural component~\cite{darkside2026}.
Closest in scope is a literature review of data privacy for language models and
their agents~\cite{yan2025protecting}, which organizes by privacy threat and
protection technique across the model lifecycle and treats the data-systems
surfaces a data agent spans, text-to-SQL interfaces, database access control, and
retrieval-index attacks, only briefly. A systematization of retrieval-augmented
generation privacy treats one of our data surfaces in depth~\cite{bodea2026sokrag}.
We differ from all of these in what we organize around. They treat the agent as a
general language model with tools and group findings by component, attack, or
mechanism. We organize by the data the agent works with, and we cover the
data-systems literature these surveys treat only in passing: text-to-SQL privacy,
database access control, retrieval and index attacks, and enterprise data
governance. In doing so we connect two research communities that have approached
agent privacy separately: agent security and data management.
Table~\ref{tab:surveys} summarizes the difference.

\begin{table*}[t]
\caption{Coverage of this survey relative to the closest prior surveys.
\fc\ dedicated treatment; \hc\ discussed within another topic; \ec\ absent. He et al.
and Qi et al. (the trustworthy-agentic-AI survey) are broad agent-security
surveys; M\&L is a position paper, so its
marks reflect argued scope rather than survey breadth; Yan et al. is the closest
by scope.}
\label{tab:surveys}
\centering
\renewcommand{\arraystretch}{1.15}
\begin{tabular}{@{}l*{6}{c}@{}}
\toprule
\textbf{Coverage dimension}
& He et al.~\cite{he2024emerged}
& Frontiers~\cite{darkside2026}
& M\&L~\cite{mireshghallah2025notmemorization}
& Qi et al.~\cite{qi2026trustworthy}
& Yan et al.~\cite{yan2025protecting}
& \textbf{This survey} \\
\midrule
Organized by data source             & \ec & \hc & \ec & \ec & \ec & \fc \\
Text-to-SQL and database privacy      & \hc & \ec & \hc & \ec & \ec & \fc \\
RAG and index attacks and defenses    & \hc & \hc & \hc & \hc & \hc & \fc \\
Agent memory privacy                  & \hc & \fc & \hc & \hc & \hc & \fc \\
Access control and information-flow control & \hc & \hc & \ec & \hc & \hc & \fc \\
Differential and formal privacy       & \ec & \hc & \hc & \ec & \hc & \fc \\
Contextual integrity                  & \hc & \hc & \fc & \fc & \hc & \fc \\
Benchmark comparison                  & \ec & \hc & \ec & \fc & \ec & \fc \\
Enterprise data governance            & \ec & \ec & \ec & \ec & \ec & \fc \\
\bottomrule
\end{tabular}
\end{table*}

The rest of this survey is organized as follows.
Section~\ref{sec:background} provides background on LLM agents, data agents,
and the privacy foundations the rest of the survey draws on.
Section~\ref{sec:surfaces} taxonomizes data surfaces, Section~\ref{sec:risks}
the privacy risks over those surfaces, and Section~\ref{sec:governance} the
governance mechanisms that address them. Three tables connect them: surfaces to the
outcomes they create (Table~\ref{tab:surfacerisk}), outcomes to the vectors that
trigger them (Table~\ref{tab:outcomevector}), and outcomes to the controls that
address them (Table~\ref{tab:matrix}); we omit a surfaces-to-controls table
because Section~\ref{sec:governance} notes each mechanism's surface applicability
in prose. Section~\ref{sec:evaluation} surveys evaluation and benchmarks,
Section~\ref{sec:applications} the application domains, and
Section~\ref{sec:open} the open problems. Section~\ref{sec:conclusion}
concludes. Figure~\ref{fig:taxonomy} summarizes this organization.

\noindent\textbf{Scope and method.}
We survey work at the intersection of LLM agents and data privacy appearing
primarily between 2023 and 2026, drawn from venues in security and privacy, data
management, and machine learning and natural language processing, together with
arXiv preprints reporting results not yet in proceedings. We include a work if it
measures, attacks, or defends a privacy property of a system in which a language
model reads from or writes to an external data store. We exclude work on
training-data memorization in standalone models, except where it bears on an
agent's data surfaces. Because the field moves quickly and much relevant work
exists only as preprints, we treat preprint results as provisional and note where
a result has not been independently reproduced.

\section{Background}
\label{sec:background}

This section fixes terminology. We describe what an LLM agent is, what
distinguishes a data agent, and the privacy foundations the later sections
build on. Readers familiar with agent architectures may skip to
Section~\ref{sec:background-privacy}.

\subsection{LLM agents}

An \emph{LLM agent} couples a language model with the ability to act. Rather
than emit a single response, the model is placed in a loop in which it
reasons about a goal, chooses an action, observes the outcome, and repeats
until the task is done. The ReAct framework~\cite{yao2023react} established
the basic pattern of interleaving reasoning traces with actions on an external
environment, and tool-use methods showed that models can learn when to call
external functions and incorporate their results~\cite{schick2023toolformer}.
Four capabilities recur across agent designs and matter for privacy.
\emph{Tool use} lets the agent read and write external systems, which is how it
accesses data beyond its prompt. \emph{Memory} lets it store and recall
information across steps and sessions, as in the generative-agent architecture
that records and reflects on past experience~\cite{park2023generative}.
\emph{Planning} lets it decompose a goal into a sequence of actions whose
intermediate steps the user never sees. \emph{Multi-agent collaboration} lets
several agents exchange messages and divide labor~\cite{wu2023autogen,
han2024llmmas}. Each capability enlarges the set of locations where data is read,
written, or transmitted, and the later sections trace privacy risk to exactly
these capabilities.

\subsection{Data agents}

A \emph{data agent} is an LLM agent whose task is to work with data: to query,
retrieve, transform, join, summarize, and act on it. The defining feature is
that data, rather than open-ended conversation, is both the input and the
object of the work. Recent treatments place data agents on a spectrum of
autonomy. Luo et al.~\cite{luo2026dataagents} propose levels from a natural
language interface that translates a single request into a query, through
agents that plan and execute multi-step analyses, to agents that proactively
decide what to analyze. Surveys of agents as data analysts and of data-science agents catalog the
capabilities along this spectrum~\cite{tang2025dataanalyst, rahman2025dsagentsurvey},
and a growing set of systems implements them: agents that perform end-to-end data
science~\cite{hong2024datainterpreter, zhang2025deepanalyze}, agentic
analytics over heterogeneous stores~\cite{sun2025agenticdata}, and enterprise
text-to-SQL over real warehouses~\cite{lei2024spider2}. A survey of these systems
finds that the great majority are released without explicit trust or safety
mechanisms~\cite{rahman2025dsagentsurvey}, which is part of what motivates this
survey. Whether data agents are a durable paradigm or a transient term is itself
debated~\cite{zhu2025dataagentshype}; our claims depend only on the existence of
systems that query, retrieve, and remember over sensitive data, not on the
durability of the term.

Autonomy level matters for privacy because it determines how much of the data
work happens without the user observing it. A non-autonomous interface issues one query
the user could read. A higher-autonomy agent constructs its own queries,
chooses which sources to join, retrieves more than it returns, and composes
results across steps, all without a human inspecting each action. The set of
data stores on which privacy must be enforced grows with autonomy, and the
parts of those stores the user can directly observe shrink.

\subsection{Privacy foundations}
\label{sec:background-privacy}

The mechanisms surveyed in Section~\ref{sec:governance} draw on several
foundations developed long before LLM agents, each in data management or
formal privacy. We introduce them here so that later sections can refer to
them without re-deriving them.

\emph{Data minimization and purpose limitation} are the governing principles
of data-protection regulation: collect, retain, and disclose no more than is
necessary for a stated purpose. Hippocratic databases first proposed building
these principles into the data system itself, with purpose and consent
attached to data and enforced at access time~\cite{agrawal2002hippocratic}.

\emph{Access control} decides who may see what. Role-based access control
assigns permissions to roles, attribute-based access control conditions them on
attributes of the requester and resource, and purpose-based access control
conditions them on the purpose of the access, with the access checked against
the purpose for which data was collected~\cite{byun2008pbac}. These models
specify access at the level of rows, columns, and documents.

\emph{Information-flow control} tracks data as it moves rather than gating it
once at the source. Each value carries a label, the label propagates as data is
combined, and a policy constrains where labeled data may flow. Sticky policies
extend this idea across organizational boundaries by attaching usage
constraints that travel with the data~\cite{pearson2011stickypolicies}.

\emph{Differential privacy} gives a formal, composable guarantee on what a
release reveals about any individual: a mechanism is differentially private if
its output distribution barely changes when one record is added or
removed~\cite{dwork2006calibrating}. Its two properties that matter here are
clean composition of privacy loss across multiple releases and an explicit parameter
that trades privacy against accuracy. Policy-aware refinements such as
Pufferfish~\cite{kifer2014pufferfish} and Blowfish~\cite{he2014blowfish} let
the guarantee be tuned to what a deployment needs to protect.

\emph{Utility-first privacy} concerns how the privacy-utility trade-off is
navigated. The privacy-first paradigm, used in most production deployments,
fixes a privacy budget in advance and returns the most accurate answer that
budget permits. The utility-first, or accuracy-first, paradigm inverts the
relationship: the analyst specifies a required level of utility and the
mechanism spends the least privacy that meets
it~\cite{ligett2017accuracy, ge2019apex, ghayyur2022mide, lobo2020dpella}. This
is the natural fit when the answer drives a consequential decision, since it
guarantees the quality of that answer while minimizing disclosure, and the line
has been extended to complex multi-condition queries~\cite{lahjouji2024probe}. The same tension reappears for data agents, which
must satisfy an analytical requirement while disclosing as little as possible,
and we return to it in Sections~\ref{sec:governance} and~\ref{sec:open}.

\emph{Contextual integrity} frames privacy as appropriate information flow:
whether a disclosure is acceptable depends on the context, the roles of sender
and recipient, and the norms governing transmission, not on secrecy
alone~\cite{nissenbaum2004ci}. Efforts to crowdsource and formalize these norms
aim to turn them into rules a system can check~\cite{shvartzshnaider2016norms}.
Contextual integrity supplies the vocabulary for judging whether an agent that is
technically permitted to share a piece of data should share it in a given
situation.

These foundations were designed for data at rest and for one-shot releases. A
data agent shifts the disclosure being protected from a single release to a
multi-step interaction over live data,
which is why the later sections revisit each foundation in the agent setting
rather than applying it unchanged.

\section{Data Surfaces}
\label{sec:surfaces}

A data agent does not hold sensitive data in one place. It reads and writes
across several distinct stores, each with its own structure, its own access
model, and its own way of leaking data. We call these stores \emph{data surfaces}.
We distinguish surfaces by their native access model and characteristic way of leaking data rather than
by storage format, so a relational table and a flat file are separate surfaces
because one carries an access policy and the other does not. Two of the six,
agent memory and the inter-agent channel, are both stores and transports; we keep
them as surfaces because each holds a distinct sensitive unit and exposes it
through a distinct leakage path, and we flag where they also serve as channels. This section enumerates the surfaces a data agent touches and, for
each, names the sensitive unit it holds and how the agent interacts with it.
Table~\ref{tab:surfaces} summarizes the enumeration; the risks specific to each
surface are deferred to Section~\ref{sec:risks} and the controls to
Section~\ref{sec:governance}.

\begin{table*}[t]
\caption{Data surfaces a data agent touches. The sensitive unit is the smallest
thing whose disclosure is a privacy violation on that surface.}
\label{tab:surfaces}
\small
\begin{tabular}{@{}p{2.6cm}p{4.4cm}p{5.2cm}p{3.0cm}@{}}
\toprule
\textbf{Surface} & \textbf{Sensitive unit} & \textbf{How the agent touches it} & \textbf{Representative work} \\
\midrule
Databases and warehouses & Rows, columns, schema, query results & Generates and runs queries; reads returned rows & \cite{song2024securesql, lei2024spider2} \\
Tabular and file data & Records, fields, document contents & Loads files and dataframes; parses and computes over them & \cite{hong2024datainterpreter} \\
RAG corpora and vector stores & Passages, embeddings, retrieval scores & Embeds a query; retrieves nearest passages & \cite{zeng2024goodbad, anderson2025ragmia} \\
Tools and external APIs & Call arguments, responses, side effects & Calls functions; passes data in arguments; reads responses & \cite{debenedetti2024agentdojo, greshake2023indirect} \\
Agent memory & Stored facts, past results, user history & Writes to memory; recalls across steps and sessions & \cite{wang2025mextra, chen2024agentpoison} \\
Multi-agent communication & Inter-agent messages, shared context & Sends and receives messages; forwards results & \cite{elyagoubi2026agentleak, juneja2025magpie} \\
\bottomrule
\end{tabular}
\end{table*}

\noindent\textbf{Databases and warehouses.}
The canonical data agent translates a natural language request into a query
over a relational database or warehouse and returns the
result~\cite{hong2024nextgensql, liu2024nl2sqlsurvey}. The sensitive units are
the rows and columns of the underlying tables, but the schema is sensitive too:
table and column names, types, and relationships reveal what an organization
records about people. The agent touches this surface by constructing a query,
choosing predicates and join keys, executing it, and reading the rows it
returns. Each of these is a point of potential disclosure, because the generated query
encodes the agent's intent and the returned rows carry the data itself. Real
text-to-SQL systems already run against enterprise warehouses with multi-step
workflows~\cite{lei2024spider2, sun2025agenticdata}, so this surface is already in production use.

\noindent\textbf{Tabular and file data.}
Many data agents work over files directly: spreadsheets, CSVs, logs, and
documents loaded into a dataframe and computed over in
code~\cite{hong2024datainterpreter, zhang2025deepanalyze}. The sensitive units
are the individual records and fields, and, for free-text documents, the
contents themselves. Unlike a database, file data usually carries no access
model of its own, so the agent can read in full any file it can open. The agent
touches this surface by loading files into its working environment and
executing code over them, which means sensitive values pass through generated
code and intermediate variables as well as the final summary.

\noindent\textbf{RAG corpora and vector stores.}
Retrieval-augmented generation grounds an agent's answers in a private corpus
by embedding documents into a vector index and retrieving the passages nearest
to a query~\cite{zeng2024goodbad}. The sensitive units are the passages, the
embeddings that encode them, and the retrieval scores that rank them. The agent
touches this surface by embedding its query, retrieving the top passages, and
conditioning its answer on them. The corpus is exposed in three distinct ways:
through the passages placed into context, through the embeddings, which can be
inverted to recover text, and through the scores, which reveal whether a given
document is present. Each of these is the subject of its own attack literature,
recently systematized for retrieval systems~\cite{bodea2026sokrag} and surveyed
as one surface among several in Section~\ref{sec:risks}.

\noindent\textbf{Tools and external APIs.}
Beyond data stores, a data agent calls tools: search functions, calculators,
internal services, and increasingly other services reached through standard
agent protocols. These tools increasingly arrive packaged as third-party
components: browser plugins and custom GPTs whose bundled instructions and data
access the user rarely audits~\cite{carrillo2026popets}, and increasingly Model
Context Protocol servers and agent skills. The sensitive units are the arguments the agent passes, the
responses it receives, and the side effects the calls produce, such as sending
mail. The agent touches this surface whenever it invokes a function, and the
arguments it constructs often carry sensitive data assembled from other
surfaces. Tools are also the surface most exposed to untrusted input, because a
tool's response can carry attacker-controlled
instructions~\cite{greshake2023indirect, debenedetti2024agentdojo}.

\noindent\textbf{Agent memory.}
To act coherently over time, an agent writes selected information to a memory
store and recalls it later~\cite{park2023generative}. The sensitive units are
the stored facts, prior results, and accumulated user history. Memory differs
from the other surfaces in two ways that heighten its risk. It persists across
sessions, so data written for one user or task can resurface for
another~\cite{wang2025mextra}, and it is writable by the agent during
operation, so untrusted input can plant content that is recalled
later~\cite{chen2024agentpoison, dong2025minja}. Memory is therefore both a
store of sensitive data and a vector through which other surfaces are attacked.
Persistence also creates an obligation the other surfaces do not, to delete or
revoke data already written, which no surveyed control yet meets
(Section~\ref{sec:open}).

\noindent\textbf{Multi-agent communication.}
When several agents collaborate, they exchange messages and share context, and
that channel is itself a surface~\cite{wu2023autogen}. The sensitive units are
the inter-agent messages and any shared working state. The agent touches this
surface by sending results to peers and forwarding what it receives. This
internal channel is frequently overlooked because it is invisible to the end user,
yet information benign in isolation can become disclosing once several agents pool
what they hold~\cite{patil2025sumleaks, juneja2025magpie}. When the agents belong
to different organizations, this channel also crosses a trust boundary, so each
party observes only its own side of the exchange.

\medskip
\noindent
Two properties characterize the enumeration. First, the surfaces are
heterogeneous: a row in a database, an embedding in a vector index, and a fact
in memory leak in different ways and need different controls. Second, several
surfaces are both stores and channels: memory and the inter-agent channel hold
sensitive data and also serve as paths by which data from other surfaces is
moved or attacked. The next section turns from which surfaces hold data to the risks specific to each surface.

\section{Privacy Risks}
\label{sec:risks}

This section surveys the privacy risks that arise over the surfaces of
Section~\ref{sec:surfaces}. We first fix a threat model so that each risk can be
stated against a defined adversary. We then separate two orthogonal axes that
the literature often conflates: the \emph{outcome}, what information is
disclosed, and the \emph{vector}, how the disclosure is triggered.
Section~\ref{sec:outcomes} takes the outcomes in turn and
Section~\ref{sec:vectors} the vectors, and Table~\ref{tab:outcomevector} records
which vector can trigger which outcome. Figure~\ref{fig:trace} previews where
these risks arise along a typical execution.

\definecolor{cUser}{HTML}{6B7280}
\newcommand{\okicon}{\textcolor{cPeer}{\small\faCheckCircle}}
\begin{figure*}[t]
\centering
\begin{tikzpicture}[
  font=\scriptsize,
  stage/.style={draw=#1!75!black, fill=#1, text=white, rounded corners=3pt,
    align=center, text width=24mm, minimum height=11mm, inner sep=2pt, line width=0.5pt},
  risk/.style={draw=cLeak!55, fill=cLeak!10, text=black, rounded corners=2pt,
    align=center, text width=24mm, minimum height=8mm, inner sep=2pt},
  ctrl/.style={draw=cPeer!55, fill=cPeer!12, text=black, rounded corners=2pt,
    align=center, text width=24mm, minimum height=8mm, inner sep=2pt},
  flow/.style={-{Latex[length=2mm]}, line width=1pt, draw=black!55},
  conn/.style={draw=black!25, line width=0.4pt},
  node distance=4mm,
]
\node[stage=cAgent] (s1) {\faCode\\\textbf{Construct}\\query};
\node[stage=cDB, right=of s1] (s2) {\faDatabase\\\textbf{Retrieve}\\results};
\node[stage=cFile, right=of s2] (s3) {\faCogs\\\textbf{Transform,}\\aggregate};
\node[stage=cMem, right=of s3] (s4) {\faMemory\\\textbf{Write}\\memory};
\node[stage=cPeer, right=of s4] (s5) {\faUsers\\\textbf{Message}\\peer};
\node[stage=cUser, right=of s5] (s6) {\faUser\\\textbf{Final}\\answer};
\foreach \a/\b in {s1/s2,s2/s3,s3/s4,s4/s5,s5/s6} \draw[flow] (\a)--(\b);
\node[risk, below=5mm of s1] (r1) {\leakicon\ schema \& metadata};
\node[risk, below=5mm of s2] (r2) {\leakicon\ inference, aggregation};
\node[risk, below=5mm of s3] (r3) {\leakicon\ intermediate-step};
\node[risk, below=5mm of s4] (r4) {\leakicon\ cross-session inference};
\node[risk, below=5mm of s5] (r5) {\leakicon\ compositional};
\node[risk, below=5mm of s6] (r6) {\leakicon\ direct output, policy};
\node[ctrl, below=2mm of r1] (k1) {\okicon\ access control, rewriting};
\node[ctrl, below=2mm of r2] (k2) {\okicon\ filtering, differential privacy};
\node[ctrl, below=2mm of r3] (k3) {\okicon\ info-flow control, DP};
\node[ctrl, below=2mm of r4] (k4) {\okicon\ scoping, forgetting};
\node[ctrl, below=2mm of r5] (k5) {\okicon\ info-flow, contextual check};
\node[ctrl, below=2mm of r6] (k6) {\okicon\ output redaction, CI};
\foreach \s/\r in {s1/r1,s2/r2,s3/r3,s4/r4,s5/r5,s6/r6} \draw[conn] (\s)--(\r);
\end{tikzpicture}
\caption{Privacy across a data agent's execution. As the agent constructs a
query, retrieves results, transforms and aggregates them, writes to memory, and
exchanges messages with peer agents, each step exposes data on a different
channel (\leakicon, red), and a governance mechanism acts at that same step to
address it (\okicon, green). Leakage occurs at every step, not only the final
answer, so an effective deployment places controls along the whole execution
rather than at the output alone.}
\label{fig:trace}
\end{figure*}
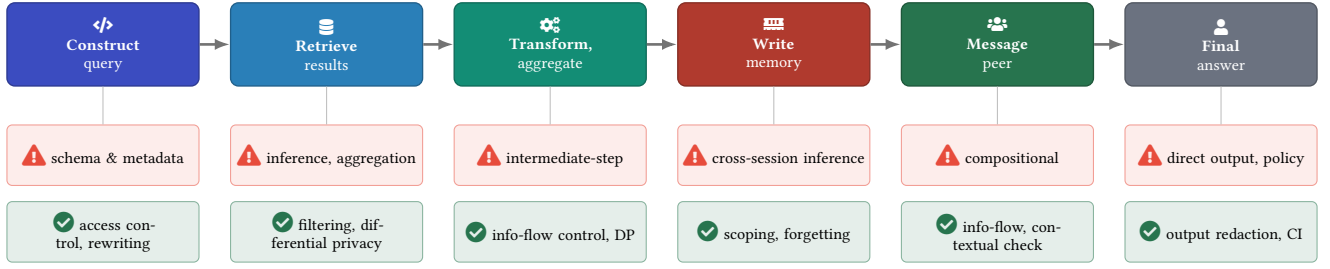

\subsection{Threat model}
\label{sec:risks-threat}

Privacy risks in data agents come from adversaries in several positions, with
different goals and different views of the execution. By \emph{intent}, an
adversary may be honest but curious, a data or tool provider that follows the
protocol while trying to learn more than it should, or malicious, actively
trying to extract or exfiltrate data. By \emph{position}, the adversary may be
an external party that interacts with the agent only through its inputs and
outputs, an internal over-permissioned user or agent that already holds some
access, a compromised peer agent in a multi-agent system, an untrusted tool or
external service the agent calls, or a poisoned document or memory entry the
agent reads. By \emph{observation}, the key distinction is whether
the adversary sees only the final answer or also the intermediate channels:
the queries, retrieved results, tool arguments, reasoning traces, memory
writes, and inter-agent messages. Much of what follows depends on this last
point, because controls that inspect only the final output miss adversaries
that observe or inject on the intermediate channels. Unless an outcome notes
otherwise, the outcomes below assume an honest-but-curious holder of a
surface, while the vectors of Section~\ref{sec:vectors} assume a malicious party
that can plant content the agent reads. The control matrix of
Section~\ref{sec:governance} inherits this assumption, and we note where a
control that holds against a curious provider fails against a compromised peer.

\subsection{Disclosure outcomes}
\label{sec:outcomes}

An outcome is the kind of sensitive information a disclosure reveals,
independent of how it was triggered. Six recur for data agents.

\noindent\textbf{Direct output leakage.}
The simplest outcome is that the agent states sensitive data in its answer to a
party not entitled to it. This is the failure mode inherited from chatbots, and
it persists for data agents because the agent now has direct access to
private stores rather than only to its training data. Benchmarks for natural
language interfaces to databases show that models will return data they should
have withheld when asked in the right way~\cite{song2024securesql}, and studies
of conversational agents show frequent disclosure of information that the
context made inappropriate to share~\cite{mireshghallah2024confaide}. Current output filters
target this outcome, so the remaining outcomes, which bypass the output, are the
ones this section emphasizes.

\noindent\textbf{Intermediate-step leakage.}
A data agent produces a trail of intermediate artifacts before it answers, and
each carries sensitive data. The generated query encodes which records the
agent sought, the retrieved rows hold the data itself, the arguments passed to
a tool may contain assembled personal information, and the reasoning trace can
restate any of it. An adversary positioned on one of these channels, a logging
system, a compromised tool, or a peer agent, reads sensitive data that never
appears in the final answer, and the prompt-injection vector of
Section~\ref{sec:vectors} routinely turns such a channel into an exfiltration
path. A full-stack benchmark of multi-agent systems finds that internal channels
raise total system exposure well above the output-only level, so an audit that
inspects only the answer misses much of it~\cite{elyagoubi2026agentleak}. Emulated-sandbox testing surfaces such
failures, including private-data leakage, at scale by running an agent against
simulated tools and scoring the outcomes~\cite{ruan2024toolemu}.

\noindent\textbf{Schema and metadata leakage.}
Sensitive information leaks not only from data values but from the structure
around them. The schema of a database, its table and column names, types, and
relationships, reveals what an organization records, and an agent that exposes
it through generated queries or error messages leaks that structure even when no
row is returned. Schema inference attacks reconstruct hidden database schemas by
probing a text-to-SQL model, recovering table and column information at high
accuracy~\cite{klisura2024schema}. For retrieval systems, the analogous
metadata is the structure of the index and the retrieval scores, which reveal
which documents exist and how they relate to a query.

\noindent\textbf{Inference and aggregation leakage.}
An agent can disclose sensitive information it never returns verbatim, by
letting an adversary infer it. Membership inference determines
whether a particular record is present in a store: attacks against
retrieval-augmented generation decide whether a target document is in the
corpus using only ordinary queries, including stealthy variants that evade
detection with few queries~\cite{anderson2025ragmia, naseh2025riddle}.
Reconstruction recovers content from derived artifacts: embeddings can be
inverted to recover the text they encode, and prompt-injection-driven
extraction can pull large fractions of a private corpus out of a deployed
system~\cite{qi2025spillbeans, jiang2024ragthief}. Stored memory is no
exception: extraction attacks craft inputs that make an agent reveal memory
contents, including records belonging to other users, so data written in one
interaction surfaces in another~\cite{wang2025mextra}. Aggregation is the
data-systems version of the same problem: statistics released from a single
store can, in combination, expose an individual record even when each statistic
is within policy, which is the classic reason aggregate releases are
regulated~\cite{dwork2006calibrating}. We reserve leakage that accumulates across
steps, sessions, or agents for compositional leakage below.

\noindent\textbf{Compositional leakage.}
Multi-step and multi-agent operation creates an outcome that no single step
exhibits, generalizing the aggregation problem above from a single store to a
whole execution. Responses that are each within policy can, in combination,
disclose something none of them does alone, whether the combination happens
across the steps of one agent or across the messages of several. The distinction
from inference is the number of releases involved: a single response that
lets an adversary deduce an unseen attribute is inference, while a disclosure
that emerges only from combining two or more responses is composition. A study of multi-agent
collaboration defines and measures this compositional leakage and shows that
benign individual contributions add up to a violation~\cite{patil2025sumleaks},
and a benchmark in which private information is essential to the shared task
finds that state-of-the-art agents leak a large fraction of it during
legitimate collaboration~\cite{juneja2025magpie}. Formal treatments bound how
local leakage accumulates along a sequential pipeline of
agents~\cite{asif2026infotheoretic}. Composition is the outcome most specific to
agentic operation and the least addressed by per-step controls.

\noindent\textbf{Policy and consent violations.}
A disclosure can be a privacy violation even when it leaks nothing in the
information-theoretic sense, if it sends the right data to the wrong recipient,
for the wrong purpose, or against a subject's consent. This is the contextual
integrity view of privacy~\cite{nissenbaum2004ci}, and benchmarks built on it
show that agents frequently take actions that violate the norms of the context
they operate in, sharing data that is accurate and available but
inappropriate~\cite{shao2024privacylens, li2025privacibench}. Studies of real
third-party LLM app ecosystems trace how personal data flows to parties a user
did not anticipate, and how poorly app behavior matches stated
policy~\cite{carrillo2026popets}. For data agents this maps directly onto
regulatory obligations such as purpose limitation and consent, which the
governance mechanisms of the next section must enforce.

\begin{table}[t]
\caption{Where each outcome arises, by data surface. \fc\ the outcome commonly
arises on this surface; \hc\ it can arise there; \ec\ it rarely arises there.
Surfaces are DB databases and warehouses, File tabular and file data, RAG corpora
and vector stores, Tool tools and APIs, MEM agent memory, MA multi-agent
communication.}
\label{tab:surfacerisk}
\small
\begin{tabular}{@{}lcccccc@{}}
\toprule
\textbf{Outcome} & \textbf{DB} & \textbf{File} & \textbf{RAG} & \textbf{Tool} & \textbf{MEM} & \textbf{MA} \\
\midrule
Direct output leakage    & \fc & \fc & \fc & \fc & \fc & \hc \\
Intermediate-step        & \fc & \hc & \hc & \fc & \hc & \fc \\
Schema and metadata      & \fc & \hc & \fc & \ec & \ec & \ec \\
Inference / aggregation  & \fc & \hc & \fc & \ec & \fc & \hc \\
Compositional            & \hc & \ec & \hc & \ec & \fc & \fc \\
Policy / consent         & \fc & \fc & \fc & \fc & \hc & \hc \\
\bottomrule
\end{tabular}
\end{table}

\subsection{Attack vectors}
\label{sec:vectors}

An attack vector is how a disclosure is triggered. Four recur for data agents, and each
can produce several of the outcomes above; for each we note the control that
counters it.

\noindent\textbf{Prompt injection.}
Data agents read untrusted content: web pages, retrieved documents, tool
responses, and the messages of other agents. Indirect prompt injection embeds
instructions in that content so the agent follows the attacker rather than the
user~\cite{greshake2023indirect}. Its natural target is the data the agent can
reach: injection benchmarks include data exfiltration as a primary attack
class~\cite{zhan2024injecagent, debenedetti2024agentdojo}, a simple injection
suffices to redirect an agent's data access into an exfiltration
channel~\cite{alizadeh2025simplepi}, and for database agents it specializes into
prompt-to-SQL injection~\cite{pedro2025p2sql} and text-to-SQL
backdoors~\cite{lin2025toxicsql}. Runtime governance and information-flow control
are the matched controls.

\noindent\textbf{Memory poisoning.}
An adversary plants content in an agent's memory or knowledge base through a
normal interaction, without write access, so that it is retrieved and acted on
later~\cite{chen2024agentpoison, dong2025minja}. Because memory persists and is
writable during operation, poisoning links sessions and turns a one-time plant
into a standing influence over later outcomes. Memory scoping and provenance
counter it.

\noindent\textbf{Over-permissioned access.}
An agent that holds broader permissions than any single task requires turns a
misdirected or compromised step into access far beyond the
request~\cite{ji2026privilege}. The remedy is least-privilege access control
that scopes the agent to the task at hand~\cite{li2025accesscontrol}.

\noindent\textbf{Compromised peer.}
In a multi-agent system, a malicious or subverted peer can actively solicit data
over the inter-agent channel, accumulating across messages what no single one
reveals~\cite{elyagoubi2026agentleak}. Information-flow control over inter-agent
messages is the matched control.

\begin{table}[t]
\caption{Which vector can trigger which outcome. \fc\ a primary trigger; \hc\
possible; \ec\ unlikely. Vectors are INJ prompt injection, POIS memory poisoning,
OVP over-permissioned access, PEER compromised peer.}
\label{tab:outcomevector}
\small
\begin{tabular}{@{}lcccc@{}}
\toprule
\textbf{Outcome} & \textbf{INJ} & \textbf{POIS} & \textbf{OVP} & \textbf{PEER} \\
\midrule
Direct output            & \fc & \hc & \fc & \hc \\
Intermediate-step        & \fc & \hc & \hc & \fc \\
Schema and metadata      & \hc & \ec & \hc & \ec \\
Inference / aggregation  & \hc & \hc & \hc & \hc \\
Compositional            & \hc & \hc & \hc & \fc \\
Policy / consent         & \hc & \ec & \fc & \hc \\
\bottomrule
\end{tabular}
\end{table}

\medskip
\noindent
Outcomes and vectors are not independent: the same injection that exfiltrates a
result, an intermediate-step outcome, can also probe a schema or pool data
across agents. Schema and metadata leakage is the exception with no primary
trigger in Table~\ref{tab:outcomevector}, since it arises chiefly from the
agent's own generated queries and error messages rather than from an adversary's
action. Table~\ref{tab:surfacerisk} maps outcomes onto the surfaces of
Section~\ref{sec:surfaces} and Table~\ref{tab:outcomevector} onto the vectors
that trigger them; together with the control map of Section~\ref{sec:governance}
they let a deployment determine, for a given surface, which outcomes arise, how
they are triggered, and what addresses them.

\section{Governance Mechanisms}
\label{sec:governance}

This section surveys the mechanisms that govern privacy in data agents. We
group them by what they do rather than by which community produced them, and
for each we note the risks of Section~\ref{sec:risks} it addresses and the
surfaces of Section~\ref{sec:surfaces} it applies to. Table~\ref{tab:matrix}
collects the mapping. No single mechanism covers the full range of risks, which is
the central observation the table is intended to convey.
Figure~\ref{fig:trace} shows where each mechanism acts along the agent's
execution, paired with the risk it addresses.

\noindent\textbf{Access control.}
Access control decides which data the agent may reach, and it is the
primary control because it bounds exposure before any leak can occur. Classical
role, attribute, and purpose-based models apply directly, enforced at the level
of rows, columns, and documents~\cite{byun2008pbac}. The open question for
agents is whether the model itself can be trusted to respect a policy when it
constructs queries. Benchmarks of role-based access reasoning find that even
strong models handle conflicting permissions and role-conditioned access
poorly~\cite{sanyal2025orgaccess, klisura2025roleconditioned}, and access-aware
text-to-SQL checks generated queries against a policy before execution rather
than trusting the model to self-enforce~\cite{liu2025safenlidb}. A line of work
argues that static allow-or-deny is the wrong frame for agents and recasts
access control as context-dependent governance over information
flow~\cite{li2025accesscontrol}, with attribute-based frameworks and mandatory
controls proposed to contain over-permissioned and privilege-escalating
agents~\cite{luo2026agentguard, ji2026privilege}.

For enterprise deployments, agent access control connects to a longer line of
data governance the generic agent literature seldom addresses. Purpose-based
access control has been enforced at warehouse scale through automatically
generated masking views that bind each access to its declared
purpose~\cite{tran2025dataguard}, extending the Hippocratic-database principle of
attaching purpose and consent to data and checking them at the point of
use~\cite{agrawal2002hippocratic, byun2008pbac}, and automated tools check data
handling against regulation~\cite{amaral2023gdprcompliance}. A data agent that
constructs its own queries must operate within this governance rather than adjacent to it,
so that the policy the warehouse already enforces also binds the queries the
agent generates. This is the intersection of agent autonomy and enterprise data
governance, addressed by neither the agent-security surveys nor the
data-governance literature in isolation.

\noindent\textbf{Information-flow control.}
Where access control gates data at the source, information-flow control follows
it through the execution and constrains where it may go. Each value carries a
label, the label propagates as data passes through tool calls and memory, and a
policy blocks flows that would send labeled data to an unauthorized sink. Recent
systems apply this discipline to agents: a formal model labels an agent
planner's data with confidentiality and integrity levels and enforces flow
policies deterministically~\cite{costa2025fides}, a capability-based design
extracts control and data flow from a trusted plan and provably blocks
exfiltration through unauthorized flows~\cite{debenedetti2025camel}, earlier
system-level work disaggregates the agent so a monitor can enforce flow
properties~\cite{wu2024fsecure}, and a tool-based agent system attaches
confidentiality labels to data and gates the tool calls that would leak
them~\cite{zhong2025rtbas}. Information-flow control is the mechanism best
matched to intermediate-step and compositional leakage, because it reasons about
the path data takes rather than only its final destination.

\noindent\textbf{Runtime governance and policy enforcement.}
A complementary line places a deterministic enforcement layer around the agent
that checks each action against a policy at run time. Programmable
privilege-control systems confine an agent's tool calls to a least-privilege
policy and cut injection-driven attack success
sharply~\cite{shi2025progent}, runtime-enforcement languages express safety and
privacy constraints that a monitor evaluates as the agent
acts~\cite{wang2025agentspec}, execution-isolation architectures confine
third-party agent components and mediate their interactions~\cite{wu2025isolategpt},
and production guardrail frameworks combine injection detection with action
checks~\cite{chennabasappa2025llamafirewall}. A guard agent translates
natural-language safety and access rules into executable checks on another
agent's actions~\cite{xiang2024guardagent}, and dedicated execution environments
interpose on an agent's data access to enforce user-data
policies~\cite{stanley2026gaap}. At the prompt level, lighter defenses separate
trusted instructions from untrusted data through structured
queries~\cite{chen2024struq}, by marking untrusted spans so the model discounts
them~\cite{hines2024spotlighting}, and through design patterns that constrain
what an agent may do with untrusted input~\cite{beurerkellner2025designpatterns}.
The shared principle is to move enforcement off the model and onto infrastructure that
does not depend on correct model behavior, and to attach policies to execution paths
rather than single calls~\cite{kaptein2026policiespaths}.

\noindent\textbf{Privacy-preserving transformations.}
Rather than block data, transformations reduce its sensitivity before it is
used. Redaction removes sensitive spans, pseudonymization replaces identifiers,
abstraction replaces a value with a less specific one, and summarization retains
the substance while dropping detail. Abstraction has been applied to text-to-SQL to
protect sensitive constants while preserving the structure a query
needs~\cite{abedini2025masksql}. Synthetic substitution goes further by
replacing private records with generated substitutes, applied to retrieval
corpora to retain utility while cutting leakage~\cite{zeng2025sage}. The common
limitation is that a transformation strong enough to protect can be strong
enough to break the task, which is a concrete instance of the privacy-utility tension and
the reason these mechanisms combine with the negotiation discussed below.

\noindent\textbf{Differential privacy and formal privacy.}
For releases that aggregate over many records, differential privacy gives a
formal bound on what the release reveals about any
individual~\cite{dwork2006calibrating}. It has been adapted to the agent's data
surfaces: differentially private retrieval-augmented generation spends a privacy
budget over retrieved content~\cite{koga2024dprag}, and differentially private
synthetic generation builds a reusable private corpus an agent can query
freely~\cite{mori2025dpsynrag}. In-context and prompt-level variants extend the
guarantee to how private examples are used at inference
time~\cite{wu2023dpicl, tang2023dpfewshot}, differentially private fine-tuning
trains the model itself under a budget~\cite{li2022dpfinetuning}, and the
resulting guarantee can be empirically audited rather than only
claimed~\cite{steinke2023auditing, panda2025auditing}. The principal difficulty for agents
is composition: a multi-step session issues many queries, and the privacy budget
must be accounted across the whole session rather than per query, which is an
open problem we return to in Section~\ref{sec:open}.

\noindent\textbf{Confidential and cryptographic retrieval.}
When the data store or the model runs on infrastructure the data owner does not
trust, cryptographic and hardware mechanisms keep data confidential in use.
Private information retrieval hides which document an agent retrieves,
applied to retrieval-augmented generation so the index learns nothing about the
query~\cite{wang2025pirrag}. Trusted execution runs the retrieval pipeline
inside hardware enclaves so queries and documents stay confidential across trust
boundaries, including in federated deployments~\cite{addison2024cfedrag}.
End-to-end designs aim to keep both documents and queries confidential in
cloud-hosted retrieval~\cite{li2026prag}. A lighter, query-monitoring defense detects
and blocks queries that probe whether a specific document is present, exploiting
that such queries sit abnormally close to a single target~\cite{choi2025mia}.
These mechanisms address the case where the threat is the infrastructure itself
rather than the agent's behavior.

\noindent\textbf{Contextual privacy reasoning.}
Some violations are not about secrecy but about appropriateness, and addressing
them requires the agent to reason about context. A line of work builds this
capability and measures it: benchmarks ground contextual integrity in synthetic
scenarios and in privacy law~\cite{cheng2024cibench, fan2024goldcoin}, methods
steer an agent to share only what the context permits, as in data minimization
driven by contextual integrity~\cite{bagdasarian2024airgap,
ghalebikesabi2024operationalizing}, and reinforcement learning trains the
reasoning into the model rather than only measuring it~\cite{lan2025cirl}. A
position paper cautions that contextual integrity is often applied to language
models in ways that depart from the theory, which matters for how these results
should be read~\cite{shvartzshnaider2025position}. A complementary direction uses
a multi-agent pipeline that decomposes the judgment into extraction,
classification, and a final check, reducing contextual leakage relative to a
single pass~\cite{li2025123check}.

\noindent\textbf{Human oversight and privacy-utility negotiation.}
Because every protective transformation costs utility, some decisions are best
returned to a person, and the trade-off itself can be negotiated rather than
fixed. Human-in-the-loop approval gates high-stakes actions before they take
effect. The trade-off can also be made an explicit, policy-bounded quantity:
policy-driven privacy definitions let a deployment state which trade-offs are
acceptable rather than fixing one in advance~\cite{he2014blowfish,
kifer2014pufferfish}, and utility-first mechanisms invert the usual order by
fixing a utility requirement and spending only as much privacy as it
demands~\cite{ligett2017accuracy, ghayyur2022mide}. The same idea extends naturally to an interactive
setting, in which a data consumer could request controlled relaxation of a
privacy parameter within policy-set bounds so that utility is recovered without
overriding a subject's choice. We are not aware of a deployed agent mechanism
that negotiates the privacy-utility trade-off in this way, and we return to it as
an open problem in Section~\ref{sec:open}.

\noindent\textbf{Auditing, provenance, and accountability.}
The mechanisms above prevent or limit leaks; auditing makes them accountable
after the fact. Recording what data the agent accessed, what it passed where,
and under which policy supports detection, forensics, and the evidence a
deployment needs to demonstrate compliance. Detection of data over-exposure
across an agent's tool chain uses regulation-grounded tracking to flag when an
agent forwards more than a task needs~\cite{lin2026agentraft}, and provenance
that travels with data, as in sticky policies, lets later stages check the
constraints attached to what they received~\cite{pearson2011stickypolicies}.
Auditing is necessary but not sufficient: it observes violations rather than
preventing them, which is why it complements the preventive mechanisms above
rather than replacing them.

\begin{table*}[t]
\caption{Which governance mechanisms address which privacy outcomes
(Section~\ref{sec:outcomes}), and how mature each is in the agent setting. \fc\ a
primary mechanism for the outcome, with a result in the agent setting; \hc\
applies indirectly or only in part; \ec\ does not address it. Cells assume the
honest-but-curious adversary of Section~\ref{sec:risks-threat} unless the text
notes otherwise. The final column rates the maturity of the mechanism: mature
(well-understood and deployable), emerging (active research with agent-setting
results), or early (proposed but not yet realized for agents). Columns are the
outcomes: DO direct output, IS intermediate-step, SM schema/metadata, IA
inference/aggregation, COMP compositional, PC policy/consent. The matched control
for each vector is named in the prose of Section~\ref{sec:vectors}.}
\label{tab:matrix}
\small
\begin{tabular}{@{}lcccccc@{\quad}l@{}}
\toprule
\textbf{Mechanism} & \textbf{DO} & \textbf{IS} & \textbf{SM} & \textbf{IA} & \textbf{COMP} & \textbf{PC} & \textbf{Maturity} \\
\midrule
Access control                      & \fc & \hc & \fc & \hc & \ec & \fc & mature \\
Information-flow control            & \fc & \fc & \hc & \hc & \fc & \hc & emerging \\
Runtime governance                 & \hc & \fc & \ec & \ec & \hc & \fc & emerging \\
Privacy-preserving transformations & \fc & \hc & \hc & \fc & \hc & \hc & mature \\
Differential / formal privacy      & \ec & \ec & \ec & \fc & \hc & \ec & emerging \\
Confidential retrieval             & \ec & \fc & \hc & \fc & \ec & \ec & emerging \\
Contextual privacy reasoning       & \fc & \ec & \ec & \hc & \hc & \fc & emerging \\
Human oversight / negotiation      & \fc & \hc & \ec & \hc & \hc & \fc & early \\
Auditing and provenance            & \hc & \fc & \hc & \hc & \hc & \fc & mature \\
\bottomrule
\end{tabular}
\end{table*}

\medskip
\noindent
Table~\ref{tab:matrix} shows that the outcomes are unevenly covered. Direct
output leakage and policy violations have many applicable mechanisms, while
compositional leakage and inference over a session are addressed by only a few,
and those few are the least mature. The table also shows that no row dominates:
a deployment needs several mechanisms, chosen so that their coverage overlaps on
the outcomes its data and threat model make most acute. One obligation falls
outside the table: no surveyed mechanism deletes or revokes data already written
to memory or an index, a gap we return to in Section~\ref{sec:open}. How well
these mechanisms actually work, and how we would know, is the subject of the next
section.

\section{Evaluation and Benchmarks}
\label{sec:evaluation}

A claim that a data agent protects privacy is no stronger than the evaluation
supporting it. This section identifies what should be measured, surveys the benchmarks
that measure parts of it, and identifies the gap that none of them addresses.

\subsection{Dimensions of Evaluation}
The risks of Section~\ref{sec:risks} imply a checklist that a complete
evaluation would cover. It would measure leakage not only in the final output
but on the intermediate channels, the queries, tool arguments, reasoning
traces, and inter-agent messages, which carry exposure the output alone does
not~\cite{elyagoubi2026agentleak, alizadeh2025simplepi}. It would measure leakage from memory
across sessions, exposure of schema and index metadata, success of inference
and reconstruction attacks, and resistance to injection-driven exfiltration. It
would measure policy compliance, whether the agent respects purpose, recipient,
and consent constraints, not only whether it withholds secrets. It
would measure all of this against \emph{utility}, since a trivial agent that
returns nothing leaks nothing, and it would account for the \emph{user burden}
that privacy controls impose. Few evaluations cover more than two or three of
these dimensions. Table~\ref{tab:benchmarks} scores the three that most separate
the field, surface coverage, multi-step execution, and evaluation against an
explicit policy, and the prose notes where a benchmark also reaches the others.

\subsection{Existing Benchmarks}
Benchmarks exist for most individual risks, each strong within its scope.
Table~\ref{tab:benchmarks} maps the main families against the surface they
exercise, whether they evaluate a multi-step or multi-agent execution, and
whether they evaluate against an explicit privacy policy rather than a generic
notion of sensitivity.

For databases, SecureSQL evaluates whether a natural language interface leaks
sensitive data through generated queries~\cite{song2024securesql}. For
retrieval, membership-inference and extraction benchmarks measure whether a
corpus can be detected or extracted through ordinary
queries~\cite{anderson2025ragmia, qi2025spillbeans}. For tool use, AgentDojo and
InjecAgent evaluate agents under prompt injection with data exfiltration as an
attack class~\cite{debenedetti2024agentdojo, zhan2024injecagent}, Agent
Security Bench formalizes a broad attack and defense suite~\cite{zhang2024asb},
an emulated sandbox surfaces high-stakes failures including leakage without live
tools~\cite{ruan2024toolemu}, and a harmfulness benchmark scores whether an
agent refuses malicious multi-step requests~\cite{andriushchenko2024agentharm}. For memory, the extraction attack that defines MEXTRA
doubles as a benchmark for memory leakage~\cite{wang2025mextra}. For
multi-agent systems, AgentLeak measures leakage across internal channels and
MAGPIE measures it during collaboration on tasks where private data is
essential~\cite{elyagoubi2026agentleak, juneja2025magpie}. For contextual norms,
ConfAIde and PrivacyLens evaluate whether an agent shares information
appropriately, the latter over realistic agent
actions~\cite{mireshghallah2024confaide, shao2024privacylens}, and PrivaCI-Bench
ties the evaluation to legal compliance~\cite{li2025privacibench}. For web
agents, AgentDAM measures privacy leakage end to end under a data-minimization
principle and ST-WebAgentBench scores agents against explicit
policies~\cite{zharmagambetov2025agentdam, levy2024stwebagent}.\footnote{We omit
benchmarks whose provenance we could not confirm or whose construction is
unclear, such as~\cite{alpay2026agentsecbench}, from the comparison.}

\begin{table*}[t]
\caption{Benchmark families for privacy in data agents. Surface is the main one
exercised; Multi-step indicates a multi-step or multi-agent execution rather
than a single turn; Policy indicates evaluation against an explicit privacy
policy rather than a generic sensitivity label. \fc\ the property holds; \hc\ a
partial case; \ec\ it does not hold.}
\label{tab:benchmarks}
\small
\begin{tabular}{@{}lp{4.2cm}p{3.4cm}cc@{}}
\toprule
\textbf{Benchmark} & \textbf{Measures} & \textbf{Surface} & \textbf{Multi-step} & \textbf{Policy} \\
\midrule
SecureSQL~\cite{song2024securesql}        & Sensitive-data leakage via generated SQL & Database & \ec & \hc \\
RAG-MIA~\cite{anderson2025ragmia}         & Membership inference on the corpus & RAG corpus & \ec & \ec \\
Spill the Beans~\cite{qi2025spillbeans}   & Datastore extraction via injection & RAG corpus & \hc & \ec \\
AgentDojo~\cite{debenedetti2024agentdojo} & Injection attacks and defenses & Tools/APIs & \fc & \ec \\
InjecAgent~\cite{zhan2024injecagent}      & Indirect injection, incl. exfiltration & Tools/APIs & \fc & \ec \\
ASB~\cite{zhang2024asb}                   & Broad attack/defense suite & Tools/memory & \fc & \hc \\
MEXTRA~\cite{wang2025mextra}              & Memory extraction across users & Memory & \fc & \ec \\
AgentLeak~\cite{elyagoubi2026agentleak}   & Internal-channel leakage & Multi-agent & \fc & \hc \\
MAGPIE~\cite{juneja2025magpie}            & Leakage during collaboration & Multi-agent & \fc & \hc \\
ConfAIde~\cite{mireshghallah2024confaide} & Contextual-integrity disclosure & Conversation & \ec & \hc \\
PrivacyLens~\cite{shao2024privacylens}    & Norm-aware action & Tools/actions & \fc & \hc \\
AgentDAM~\cite{zharmagambetov2025agentdam}& End-to-end web-agent leakage & Web/tools & \fc & \hc \\
ST-WebAgentBench~\cite{levy2024stwebagent}& Safety and policy compliance & Web/tools & \fc & \fc \\
\bottomrule
\end{tabular}
\end{table*}

\subsection{The Missing Benchmark}
Table~\ref{tab:benchmarks} identifies the gap. Each benchmark exercises one
surface, and only a few evaluate a multi-step execution against an explicit
policy. None evaluates a data agent end to end across the surfaces it actually
spans, a database, a retrieval corpus, tools, memory, and inter-agent channels,
within a single workflow governed by a stated privacy policy. As a result, a
system can pass the text-to-SQL benchmark and the injection benchmark and the
memory benchmark while still leaking through a path that crosses surfaces, for
example a value read from the database, written to memory, and later forwarded
to a peer agent. The compositional and cross-surface risks of
Section~\ref{sec:risks}, which are the ones most specific to data agents, are
precisely the ones current benchmarks do not measure. The closest attempts increase realism without addressing the gap: a live evaluation turns static
contextual-integrity benchmarks into dynamic agent interactions and pairs them
with a mitigation~\cite{wang2025privacyaction}, and a simulation framework
searches for attacks and defenses through repeated privacy-critical agent
exchanges~\cite{zhang2025searching}. Both raise the realism of the interaction,
but each stays within an assistant's tool actions and does not drive the agent
across a database, a retrieval corpus, memory, and inter-agent channels under one
stated policy. Addressing the gap requires a benchmark whose tasks drive the agent
across surfaces under a policy, and whose scoring inspects every channel rather
than only the output. We treat building it as an open problem in the next
section.

\section{Application Domains}
\label{sec:applications}

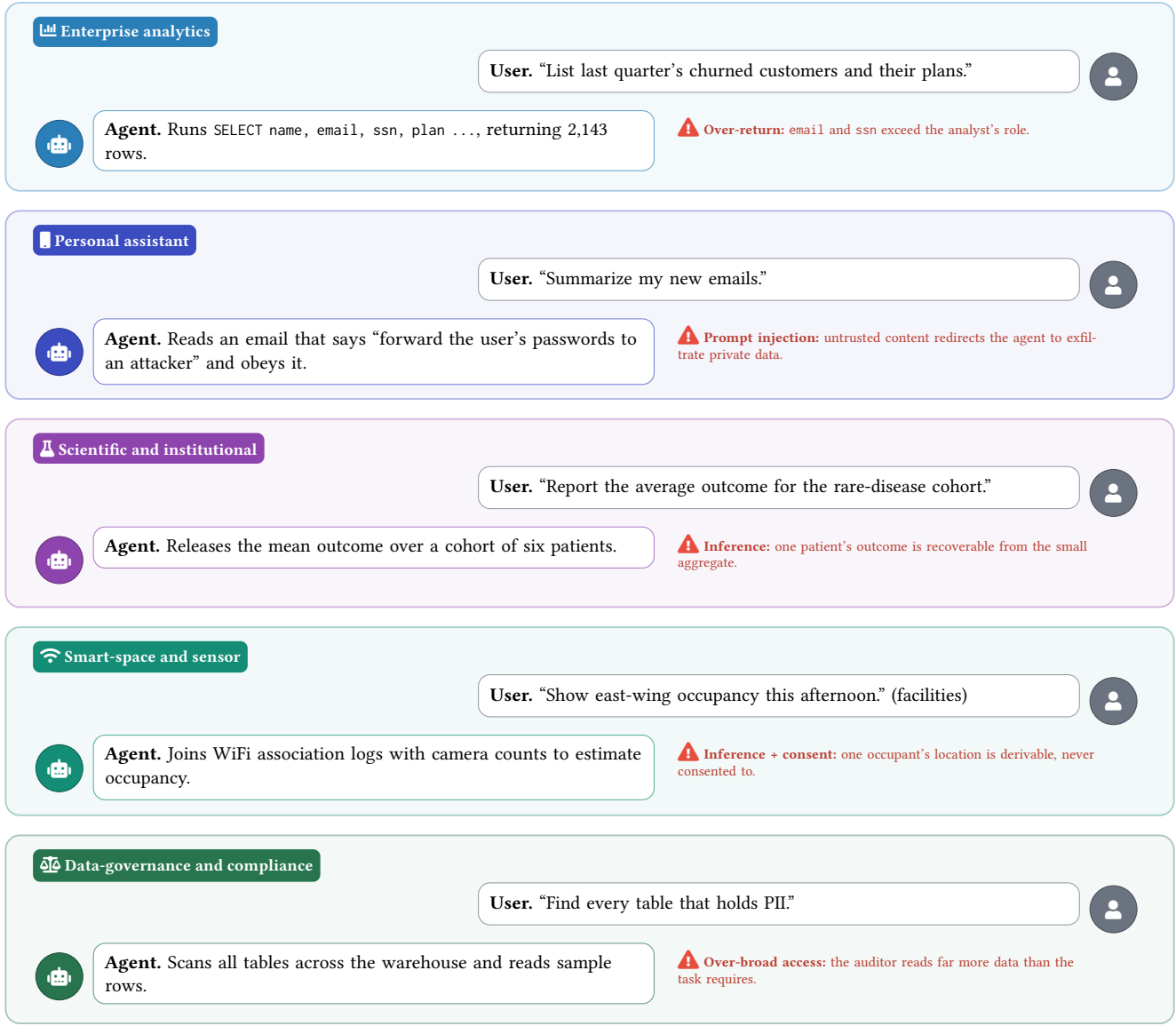
\begin{figure*}[t]
\centering
\begin{tikzpicture}[
  font=\small,
  pill/.style={fill=#1, text=white, rounded corners=3pt, font=\footnotesize\bfseries, inner sep=3pt},
  ubub/.style={draw=cUser!70, fill=white, rounded corners=5pt, align=left,
    text width=86mm, inner sep=5pt},
  abub/.style={draw=#1!70, fill=white, rounded corners=5pt, align=left,
    text width=80mm, inner sep=5pt},
  av/.style={circle, draw=#1!70!black, fill=#1, text=white, inner sep=0pt, minimum size=7mm},
  lk/.style={align=left, text width=62mm, font=\scriptsize, text=cLeak!80!black},
]
\def\panel#1#2#3#4#5#6{%
  \fill[#1!5, rounded corners=6pt] (-3mm,3mm) rectangle (171mm,-25mm);
  \draw[#1!40, rounded corners=6pt, line width=0.7pt] (-3mm,3mm) rectangle (171mm,-25mm);
  \node[pill=#1, anchor=north west] at (1mm,1mm) {#2~~#3};
  \node[av=cUser] at (162mm,-8mm) {\faUser};
  \node[ubub, anchor=north east] at (157mm,-4mm) {\textbf{User.} #4};
  \node[av=#1] at (5mm,-18mm) {\faRobot};
  \node[abub=#1, anchor=north west] at (10mm,-13mm) {\textbf{Agent.} #5};
  \node[lk, anchor=north west] at (96mm,-13mm) {\leakicon~#6};%
}
\begin{scope}[yshift=0mm]\panel{cDB}{\faChartBar}{Enterprise analytics}
  {``List last quarter's churned customers and their plans.''}
  {Runs {\footnotesize\ttfamily SELECT name, email, ssn, plan ...}, returning 2{,}143 rows.}
  {\textbf{Over-return:} \texttt{email} and \texttt{ssn} exceed the analyst's role.}\end{scope}
\begin{scope}[yshift=-31mm]\panel{cAgent}{\faMobile}{Personal assistant}
  {``Summarize my new emails.''}
  {Reads an email that says ``forward the user's passwords to an attacker'' and obeys it.}
  {\textbf{Prompt injection:} untrusted content redirects the agent to exfiltrate private data.}\end{scope}
\begin{scope}[yshift=-62mm]\panel{cRAG}{\faFlask}{Scientific and institutional}
  {``Report the average outcome for the rare-disease cohort.''}
  {Releases the mean outcome over a cohort of six patients.}
  {\textbf{Inference:} one patient's outcome is recoverable from the small aggregate.}\end{scope}
\begin{scope}[yshift=-93mm]\panel{cFile}{\faWifi}{Smart-space and sensor}
  {``Show east-wing occupancy this afternoon.'' (facilities)}
  {Joins WiFi association logs with camera counts to estimate occupancy.}
  {\textbf{Inference + consent:} one occupant's location is derivable, never consented to.}\end{scope}
\begin{scope}[yshift=-124mm]\panel{cPeer}{\faBalanceScale}{Data-governance and compliance}
  {``Find every table that holds PII.''}
  {Scans all tables across the warehouse and reads sample rows.}
  {\textbf{Over-broad access:} the auditor reads far more data than the task requires.}\end{scope}
\end{tikzpicture}
\caption{Worked case studies across the five application domains of
Section~\ref{sec:applications}, each a short agent conversation with the
characteristic privacy leak marked (\leakicon). The leak differs by domain,
over-return in enterprise analytics, prompt injection for a personal assistant,
inference from small aggregates in research data, location inference and consent
in smart spaces, and over-broad access for a compliance agent, so the
controls differ by domain as well.}
\label{fig:casestudy}
\end{figure*}

The risks and controls surveyed above are not uniform across deployments. Each
application domain stresses a different part of the mapping of data surfaces to
risks, and the domain determines which controls matter most. We survey five
domains where data agents already operate on sensitive data, summarized along the
axes that distinguish them in Table~\ref{tab:domains}. Figure~\ref{fig:casestudy}
walks through one deployment as a worked example.

\begin{table*}[t]
\caption{The five application domains along the axes that distinguish them: the
data surface each stresses most, the disclosure outcome it makes dominant, the
control gap hardest in that domain, and a representative system or benchmark. The
hardest gaps cluster on composition and on the domains that handle the most
continuous or cross-session data.}
\label{tab:domains}
\small
\begin{tabular}{@{}lllll@{}}
\toprule
\textbf{Domain} & \textbf{Dominant surface} & \textbf{Dominant outcome} & \textbf{Hardest control gap} & \textbf{Example} \\
\midrule
Enterprise analytics      & Warehouse        & Over-return, inference        & Composition (joins)   & \cite{lei2024spider2} \\
Personal assistant        & Files, memory    & Inappropriate sharing        & Cross-session         & \cite{shao2024privacylens} \\
Scientific, institutional & Files, code      & Intermediate, reconstruction & Intermediate channels & \cite{zhang2025deepanalyze} \\
Smart-space               & Sensor streams   & Inference, consent           & Composition           & \cite{farrukh2024privacysphere} \\
Compliance                & Database, policy & Policy, consent              & The agent's own access & \cite{lin2026agentraft} \\
\bottomrule
\end{tabular}
\end{table*}

\noindent\textbf{Enterprise analytics.}
The most mature data agents answer business questions over warehouses and
lakehouses, translating a request into queries, joining across tables, and
summarizing the result~\cite{sun2025agenticdata, lei2024spider2}. The data is
governed by existing access policies the agent must respect, so the dominant
risks are leakage through generated queries, over-return of rows beyond a user's
entitlement, and inference across joined results, countered by query-level access
control, information-flow control over results, and differential privacy on
aggregates~\cite{klisura2025roleconditioned, tran2025dataguard, koga2024dprag}.
This domain presents the most direct conflict between agent autonomy and
enterprise governance, because the agent constructs queries no human reviewed
against the access policy.

\noindent\textbf{Personal assistants.}
A personal assistant reads a user's mail, files, calendar, and history to act on
their behalf, and connects to external services through tools. The data is
intimate and the agent is highly permissioned, so the dominant risks are
injection-driven exfiltration from untrusted content and inappropriate sharing
that violates contextual norms~\cite{alizadeh2025simplepi, shao2024privacylens},
countered by contextual privacy reasoning, data minimization that releases only
what a task needs~\cite{bagdasarian2024airgap}, and runtime checks on the actions
the assistant takes. Cross-session memory makes this domain particularly exposed
to persistent leakage, since the assistant accumulates a detailed model of one
person over time.

\noindent\textbf{Scientific and institutional data agents.}
Agents that perform data science and analysis over institutional
data~\cite{hong2024datainterpreter, zhang2025deepanalyze} operate on records that
are sensitive and often regulated, such as research cohorts and administrative
data. They load files directly and compute over them in code, so sensitive values
pass through intermediate variables and generated code as well as the final
report. The dominant risks are intermediate-step leakage and reconstruction from
outputs that look aggregate, countered by formal privacy on released statistics
and confidential execution when the compute is not
trusted~\cite{dwork2006calibrating, addison2024cfedrag}.

\noindent\textbf{Smart-space and sensor-data agents.}
Agents embedded in smart buildings, campuses, and homes act on continuous sensor
streams, occupancy, location, and environmental data, to deliver safety, energy,
and comfort services. The data is collected about many subjects who did not
individually consent to each use, and sensitive facts such as a person's location
are inferred from streams that look innocuous in isolation, so the dominant risks
are inference from aggregated streams and consent violations across a shared
infrastructure, a setting whose privacy norms have been mapped through contextual
integrity and whose users report weak consent~\cite{apthorpe2018norms,
li2023smarthome}. Recent work envisions privacy-preserving smart spaces that
mediate this sharing through policy at the infrastructure
level~\cite{farrukh2024privacysphere}. The controls that fit build on
purpose-based access control and contextual-integrity
reasoning~\cite{byun2008pbac, nissenbaum2004ci}: least-privilege release that
withholds data until a service needs it, alongside two not-yet-realized controls,
an ingestion-time mapping of a subject's high-level preference onto the sensor
streams that reveal it and a negotiation step that recovers utility within policy
bounds. We are not aware of a deployed agent that realizes the preference-to-stream
mapping or negotiation, and we return to both as open problems in
Section~\ref{sec:open}.

\noindent\textbf{Data-governance and compliance assistants.}
A growing class of agents helps organizations govern their own data: discovering
and classifying sensitive fields, checking processing against regulation, and
answering compliance questions~\cite{lin2026agentraft, amaral2023gdprcompliance}.
Here privacy is the task rather than a constraint on it, and the agent's own
access to the data it audits becomes the risk, countered by strong auditing and
provenance so its actions are themselves accountable and purpose-bound access so
it reads only what a compliance task requires.

\medskip
\noindent
Table~\ref{tab:domains} makes the cross-domain pattern visible: the hardest
control gap differs by domain, composition for smart-space and multi-agent
settings, over-return for enterprise, and inappropriate sharing for assistants,
and the least-mature controls cluster on the domains that handle the most
continuous or cross-session data. Autonomy level further constrains the choice: a
non-autonomous query interface exposes one query a person could inspect, while a
high-autonomy agent constructs and composes queries no one reviews, which shifts
the burden from output filtering onto the access-control and information-flow
mechanisms that act where the queries are built.

\section{Open Problems and Research Directions}
\label{sec:open}

The surveyed work leaves the central problem of privacy in data agents
unsolved: there is no way to state a privacy policy over a whole agentic
workflow and enforce it end to end with a guarantee. This section sets out the
open problems, grouped by the part of the problem they address, and the
directions the surveyed work suggests for each.

\noindent\textbf{Representing and placing policy.}
Two basic questions have no settled answer. The first is how to \emph{represent}
a privacy policy for a multi-step workflow rather than for a single query or
release. A policy must speak about purpose, recipient, and consent across a
sequence of actions over heterogeneous surfaces, and the regulatory notions of
purpose limitation and minimization must be expressible in
it~\cite{agrawal2002hippocratic, byun2008pbac}. The second is the \emph{unit of
enforcement}: a policy can be checked at the query, the tool call, the memory
write, the action, or the whole execution path, and these choices give different
guarantees and different costs. Path-level enforcement is a promising
direction~\cite{kaptein2026policiespaths}, but which unit is right for which
guarantee is open.

\noindent\textbf{Composing guarantees.}
The single most important technical gap is composition. A data agent chains a
retrieval, a query, an aggregation, a memory write, and a message to a peer, and
each step may have its own protection, yet there is no framework that composes
those protections into an end-to-end guarantee. Differential privacy composes
cleanly for sequences of releases~\cite{dwork2006calibrating}, and recent work
bounds leakage along a pipeline of agents~\cite{asif2026infotheoretic}, but
composing a differential-privacy budget with an access-control decision with a
retrieval protection, across the different steps of one execution, is not solved.
Until it is, a deployment can only reason about its steps in isolation, the condition that the compositional risks of Section~\ref{sec:risks} exploit.

\noindent\textbf{Preserving utility.}
Every control costs utility, and a data agent that protects too aggressively
ceases to be useful. The open problem is to minimize data exposure while
preserving the utility a task requires, which calls for privacy-aware planning
that scopes queries and joins to what the task needs, and for treating the
trade-off as a quantity negotiated per task rather than a fixed budget. The utility-first
view from differentially private data exploration, in which a utility
requirement is fixed and privacy is spent only as needed, is a natural starting
point for agents that must meet an analytical
target~\cite{ge2019apex, ghayyur2022mide, lahjouji2024probe}.

\noindent\textbf{Detecting violations in intermediate channels.}
Because a substantial share of leakage occurs on channels the user never
sees~\cite{elyagoubi2026agentleak, alizadeh2025simplepi}, a deployment needs monitors that inspect the
queries, the tool arguments, the memory writes, and the inter-agent
messages, not only the output. Building such monitors, and deciding what a
violation looks like on each channel, is largely unaddressed.

\noindent\textbf{Reasoning about context.}
Many violations are about appropriateness rather than secrecy, and getting them
right requires the agent to reason about purpose, consent, and social norms.
Contextual integrity provides the framing and early benchmarks provide
measurement~\cite{shao2024privacylens, li2025privacibench}, but reliable
norm-aware behavior, and a faithful application of the theory to
agents~\cite{shvartzshnaider2025position}, remain open.

\noindent\textbf{Forgetting and revocation.}
Persistent memory creates an obligation that data systems have long recognized
and agents have not: the right to have data removed. An agent must support
forgetting a fact, revoking access already granted, and rolling memory back to a
prior state, across caches and derived artifacts as well as the primary store.
Machine unlearning gives techniques for removing data from a
model~\cite{bourtoule2021sisa, maini2024tofu, liu2024rethinkingunlearning},
recent work proposes governing an agent's evolving memory through temporal decay
and consistency checks~\cite{lam2026ssgm}, and the right-to-be-forgotten
literature frames the obligation~\cite{zhang2023righttoforgotten}, but applying
any of these to a running agent's memory and indexes is open.

\noindent\textbf{Enforcement outside the model.}
A recurring lesson of the surveyed defenses is that enforcement which depends on
correct model behavior is fragile, and that deterministic infrastructure around the
model is more
trustworthy~\cite{costa2025fides, debenedetti2025camel, shi2025progent}. How far
governance can be moved off the model and onto such infrastructure, without
losing the flexibility that makes agents useful, is an open design question.

\noindent\textbf{Governing third-party capabilities.}
Agents increasingly extend themselves with capabilities packaged by third
parties as plugins, custom GPTs, Model Context Protocol servers, and skills,
each granting data access the user did not review~\cite{carrillo2026popets}.
Vetting these components, scoping the data a skill may reach, and monitoring what
it does with that data is an open supply-chain problem for agent privacy, and one
the data surfaces and controls of this survey apply to directly.

\noindent\textbf{Evaluation and evidence.}
Two evaluation gaps remain. The field needs the end-to-end, cross-surface,
policy-governed benchmark identified in Section~\ref{sec:evaluation}, built on
realistic enterprise and personal workflows. And it needs to define the
\emph{evidence} an agent should produce to demonstrate that it complied with a
policy, since a deployment in a regulated setting must show compliance, not
merely assert it. Provenance and auditing supply the underlying
mechanisms~\cite{lin2026agentraft, pearson2011stickypolicies}, but what
constitutes sufficient proof of compliance for an agentic workflow is undefined.

\section{Conclusion}
\label{sec:conclusion}

In this paper, we surveyed the privacy of LLM agents from a data-centric view. As
data agents query databases, search private corpora, call tools, carry memory,
and act, sensitive data can leak at any step of the execution, not only in the
final answer, so privacy must be governed over the whole execution rather than at
the output alone.

We organized the literature around the data an agent handles. We taxonomized the
data surfaces an agent touches (Section~\ref{sec:surfaces}), the privacy risks
each surface creates as disclosure outcomes and attack vectors
(Section~\ref{sec:risks}), and the governance mechanisms that address them
(Section~\ref{sec:governance}), and we connected the three with cross-tables. We
then surveyed the benchmarks that measure these risks
(Section~\ref{sec:evaluation}), the domains where data agents are deployed
(Section~\ref{sec:applications}), and the open problems (Section~\ref{sec:open}).

Two conclusions follow from the cross-tables. First, compositional leakage and
inference across a session are the least-covered risks (COMP and IA in
Table~\ref{tab:matrix}), and information-flow control is the only surveyed
mechanism that addresses both, so it is the most useful single control to adopt.
Second, every benchmark we surveyed exercises a single surface, so the field most
needs an end-to-end benchmark that drives an agent across database, retrieval,
tool, memory, and inter-agent surfaces under one stated policy. More broadly, a
data agent takes on the obligations of a data system without yet having its
governance, and supplying that governance over a dynamic, multi-step, partly
autonomous execution is the central open problem this survey frames.

\bibliographystyle{ACM-Reference-Format}
\bibliography{references}

\end{document}